\documentclass[amsmath,amssymb,twocolumn]{revtex4}
\usepackage{graphicx}% Include figure files
\usepackage{amscd,amsmath,amsthm,amsfonts,amssymb}
\def\[{\begin{equation}}
\def\]{\end{equation}}

\begin{document}
\title{Rogue wave patterns associated with Adler-Moser polynomials in the nonlinear Schr\"odinger equation}
\author{%%%% Author details
Bo Yang$^{1}$, Jianke Yang$^{2}$}
%%%%%%%%% Insert author address here
\address{$^{1}$ School of Mathematics and Statistics, Ningbo University, Ningbo 315211, China\\
$^{2}$ Department of Mathematics and Statistics, University of Vermont, Burlington, VT 05401, U.S.A}
\begin{abstract}
We report new rogue wave patterns in the nonlinear Schr\"{o}dinger equation. These patterns include heart-shaped structures,  fan-shaped sectors, and many others, that are formed by individual Peregrine waves. They appear when multiple internal parameters in the rogue wave solutions get large. Analytically, we show that these new patterns are described asymptotically by root structures of Adler-Moser polynomials through a dilation. Since Adler-Moser polynomials are generalizations of
the Yablonskii-Vorob'ev polynomial hierarchy and contain free complex parameters, these new rogue patterns associated with Adler-Moser polynomials are much more diverse than previous rogue patterns associated with the Yablonskii-Vorob'ev polynomial hierarchy.
We also compare analytical predictions of these patterns to true solutions and demonstrate good agreement between them.
\end{abstract}
\maketitle

\section{Introduction}
Rogue waves are unusually large, mysterious, and suddenly appearing surface water waves that can be dangerous even to large ships \cite{Peli_book2009}. Their counterparts in optics and many other physical fields have also been reported \cite{Solli2007,Wabnitz2017,RomeroRos2022}. Due to their unexpected nature and potential damage, rogue waves have been heavily studied in the physical and mathematical communities in recent years. Physically, a lot of laboratory experiments on rogue waves have been performed in diverse fields such as optical fibers \cite{Kibler2010,Xu2019,Frisquet2016,Baronio2018}, water tanks \cite{Chabchoub2011,Chabchoub2012a,Chabchoub2012b,Chabchoub2013}, superfluid helium \cite{Helium2008}, plasma \cite{Bailung2011,Plasma2016}, and Bose-Einstein condensates \cite{Engles2023}. Mathematically, rogue wave studies have been greatly facilitated by the fact that, many integrable equations that govern diverse physical processes, such as the nonlinear Schr\"odinger (NLS) equation for wave packet evolution in deep water, optical fibers, plasma, and Bose-Einstein condensates \cite{Benney,Ablowitzbook, Hasegawabook, Saito1984, Gross,Pitaevskii, Kevrekidis2022}, and the Manakov system for light transmission in randomly birefringent optical fibers \cite{Bergano}, admit explicit solutions that exhibit rogue-wave characteristics. The first such solution was reported by Peregrine in the NLS equation \cite{Peregrine1983}. This Peregrine solution starts from a slightly perturbed constant-amplitude background wave. Then, it develops a localized peak that is three times the height of the background. Afterwards, this peak decays and merges to the background again. This is a rogue wave since its transient peak is much higher than its original wave amplitude, and this peak appears and disappears unexpectedly. It turns out that the Peregrine solution is just the simplest (fundamental) rogue wave in the NLS equation. More intricate NLS rogue waves that could reach even higher transient peak amplitudes or multiple peaks were later discovered \cite{AAS2009,DGKM2010,AKA2011,KAAN2011,GLML2012,OhtaJY2012,Miller2019}. In addition, rogue waves were also derived for many other integrable systems such as the Manakov system \cite{BDCW2012,ManakovDark,LingGuoZhaoCNLS2014,
Chen_Shihua2014,Chen_Shihua2015,ZhaoGuoLingCNLS2016} and so on. These analytical expressions of rogue waves shedded much light on the intricate rogue wave dynamics and guided their observations in laboratory experiments
\cite{Kibler2010,Xu2019,Frisquet2016,Baronio2018,Chabchoub2011,Chabchoub2012a,Chabchoub2012b,Chabchoub2013,Helium2008,Bailung2011,Engles2023}.

Pattern formation in rogue waves is an important question, since such information allows prediction of later rogue wave shapes from earlier rogue wave forms. One of the simplest rogue patterns is a rogue triplet, which comprises three fundamental rogue waves forming a triangle in the space-time plane. Such rogue triplets have been reported theoretically in many integrable equations \cite{DGKM2010,AKA2011,OhtaJY2012,Miller2019,Chen_Shihua2014,Chen_Shihua2015} and observed experimentally in both water tanks and randomly birefringent optical fibers \cite{Chabchoub2013,Baronio2018}. Beyond rogue triplets, more sophisticated rogue patterns comprising Peregrine waves forming shapes such as pentagons, heptagons, and rings have also been reported for the NLS equation \cite{KAAN2011,HeFokas,KAAN2013,YangNLS2021}. In addition, it was revealed in \cite{YangNLS2021} that those sophisticated patterns would reliably arise when one of the internal parameters in NLS rogue wave solutions gets large, and their shapes are predicted asymptotically by root structures of the Yablonskii-Vorob'ev polynomial hierarchy through a dilation and rotation (each nonzero root of the hierarchy predicts the spatial-temporal location of a Peregrine wave). In \cite{YangNLS2021b}, it was further revealed that those NLS rogue patterns associated with root structures of the Yablonskii-Vorob'ev polynomial hierarchy are universal and would arise in many other integrable systems, such as the derivative NLS equations, the Boussinesq equation, and the Manakov system.

Given the importance of the NLS equation in diverse physical disciplines, an important question is whether this equation admits other types of rogue patterns. If so, what special polynomials can be used to predict those patterns?

In this paper, we show that the NLS equation does admit many new rogue patterns. These patterns arise when \emph{multiple} internal parameters in rogue wave solutions get large, and their shapes are predicted asymptotically by root structures of Adler-Moser polynomials through a simple dilation. Adler-Moser polynomials are generalizations of the Yablonskii-Vorob'ev polynomial hierarchy, and their root structures are much more diverse than those of the Yablonskii-Vorob'ev polynomial hierarchy. As a consequence, we report many new rogue patterns in the NLS equation, such as heart-shaped structures, fan-shaped sectors, and others. Our analytical predictions for these new patterns based on root structures of Adler-Moser polynomials are compared to true rogue solutions, and excellent agreement is demonstrated.

\vspace{-0.3cm}
\section{Preliminaries}
The NLS equation is
\[ \label{NLS-2020}
\textrm{i} u_{t} +  \frac{1}{2}u_{xx}+ |u|^2 u=0.
\]
This equation governs nonlinear wave packet evolution in numerous physical systems such as deep water, optical fibers, plasma, and two-component Bose-Einstein condensates \cite{Benney,Ablowitzbook,Hasegawabook,Saito1984,Kevrekidis2022}.

\vspace{-0.2cm}
\subsection{General bilinear rogue wave solutions}
Rogue wave solutions in this NLS equation have been derived by various methods before \cite{DGKM2010,GLML2012,OhtaJY2012,Miller2019}.
The most explicit forms of these rogue wave solutions are the ones that were derived by the bilinear method in \cite{OhtaJY2012} and then further simplified in \cite{YangNLS2021}. These simplified bilinear rogue wave solutions of the $N$-th order are
\begin{equation}
u_N(x,t)=\frac{\sigma_{1}}{\sigma_{0}}e^{\textrm{i}t}, \label{NLSNRWs}
\end{equation}
\begin{equation} \label{sigma_n}
\sigma_{n}=
\det_{
\begin{subarray}{l}
1\leq i, j \leq N
\end{subarray}
}
\left(
\begin{array}{c}
 \phi_{2i-1,2j-1}^{(n)}
\end{array}
\right),
\end{equation}
\begin{equation}
\phi_{i,j}^{(n)}=\sum_{\nu=0}^{\min(i,j)} \frac{1}{4^{\nu}} \hspace{0.06cm} S_{i-\nu}(\textbf{\emph{x}}^{+}(n) +\nu \textbf{\emph{s}})  \hspace{0.06cm} S_{j-\nu}(\textbf{\emph{x}}^{-}(n) + \nu \textbf{\emph{s}}),
\end{equation}
\begin{equation}
x_{1}^{\pm}=x \pm \textrm{i} t \pm n, \quad x_{2k}^{\pm} = 0,
\end{equation}
\begin{equation}
x_{2k+1}^{+}= \frac{x+2^{2k} (\textrm{i} t)}{(2k+1)!} +a_{2k+1},    \quad x_{2k+1}^{-}=(x_{2k+1}^{+})^*,
\end{equation}
where $S_k(\mbox{\boldmath $x$})$ with $\emph{\textbf{x}}=\left( x_{1}, x_{2}, \ldots \right)$ are Schur polynomials defined by the generating function
\begin{equation} \label{Schurdef}
\sum_{k=0}^{\infty}S_k(\mbox{\boldmath $x$}) \epsilon^k
=\exp\left(\sum_{k=1}^{\infty}x_k \epsilon^k\right),
\end{equation}
$\textbf{\emph{s}}=(0, s_2, 0, s_4, \cdots)$ are coefficients from the expansion
\begin{equation} \label{sexpand}
\sum_{j=1}^{\infty} s_{j}\lambda^{j}=\ln \left[\frac{2}{\lambda}  \tanh \left(\frac{\lambda}{2}\right)\right],
\end{equation}
the asterisk * represents complex conjugation, and $a_3, a_5, \cdots, a_{2N-1}$ are free irreducible complex parameters which control the shape of this rogue wave solution.

When $N=1$, the above solution is $u_1(x,t)=\hat{u}_1(x, t)\hspace{0.04cm} e^{\textrm{i}t}$, where
\begin{equation} \label{Pere}
\hat{u}_1(x, t)=1- \frac{4(1+2\textrm{i}t)}{1+4x^2+4t^2}.
\end{equation}
This is the fundamental rogue wave in the NLS equation that was discovered by Peregrine in \cite{Peregrine1983}.

\vspace{-0.2cm}
\subsection{Adler-Moser polynomials and their root structures} \label{secPIII}

Adler-Moser polynomials were proposed by Adler and Moser \cite{Adler_Moser1978}, who expressed rational solutions of the
Korteweg-de Vries equation in terms of those polynomials. In a different context of point vortex dynamics, it was discovered unexpectedly that the zeros of these polynomials also form stationary vortex configurations when the vortices have the same strength but positive or negative orientations, and the numbers of those positive and negative vortices are consecutive triangular numbers \cite{Aref2007FDR,Clarkson2009}.

Adler-Moser polynomials $\Theta_{N}(z)$ can be written as a determinant \cite{Clarkson2009}
\begin{equation} \label{AdlerMoserdef}
\Theta_{N}(z) = c_{N} \left| \begin{array}{cccc}
         \theta_{1}(z) &\theta_{0}(z) & \cdots &  \theta_{2-N}(z) \\
         \theta_{3}(z) & \theta_{2}(z) & \cdots &  \theta_{4-N}(z) \\
        \vdots& \vdots & \vdots & \vdots \\
         \theta_{2N-1}(z) & \theta_{2N-2}(z) & \cdots &  \theta_{N}(z)
       \end{array} \right|,
\end{equation}
where $\theta_{k}(z)$ are Schur polynomials defined by
\begin{equation}\label{AMthetak}
\sum_{k=0}^{\infty} \theta_k(z) \epsilon^k =\exp\left( z \epsilon + \sum_{j=1}^{\infty} \kappa_j \epsilon^{2j+1} \right),
\end{equation}
$\theta_{k}(z)\equiv 0$ if $k<0$, $c_{N}= \prod_{j=1}^{N}(2j-1)!!$, and $\kappa_j \hspace{0.04cm} (j\ge 1)$ are arbitrary complex constants. Note that our $\kappa_j$ constant is slightly different from that in \cite{Clarkson2009} by a factor of $-1/(2j+1)$, and this different parameter definition will be more convenient for our purpose. The determinant in (\ref{AdlerMoserdef}) is a Wronskian since we can see from Eq.~(\ref{AMthetak}) that $\theta'_{k}(z)=\theta_{k-1}(z)$, where the prime denotes differentiation. In addition, these $\Theta_{N}(z)$ polynomials are monic with degree $N(N+1)/2$, which can be seen by noticing that the highest $z$ term of $\theta_k(z)$ is $z^k/k!$, and the determinant in (\ref{AdlerMoserdef}) with $\theta_k(z)$ replaced by its highest $z$ term can be explicitly calculated as $z^{N(N+1)/2}$ \cite{OhtaJY2012}. Adler-Moser polynomials reduce to the Yablonskii-Vorob'ev polynomial hierarchy when all $\kappa_j$ constants are set as zero except for one of them \cite{YangNLS2021}. Thus, we can view Adler-Moser polynomials as generalizations of the Yablonskii-Vorob'ev polynomial hierarchy.

The first few Adler-Moser polynomials are
\begin{eqnarray*}
&& \hspace{-0.8cm} \Theta_1(z)=z, \\
&& \hspace{-0.8cm} \Theta_2(z)=z^3-3 \kappa _1, \\
&& \hspace{-0.8cm} \Theta_3(z)=z^6-15 \kappa _1 z^3+45 \kappa _2 z-45 \kappa _1^2,  \\
&& \hspace{-0.8cm} \Theta_4(z)=z^{10}-45 \kappa _1 z^7+315 \kappa _2 z^5-1575 \kappa _3 z^3  \\
&& \hspace{0.0cm} +4725 \kappa _1 \kappa _2 z^2-4725 \kappa _1^3 z-4725 \kappa _2^2+4725 \kappa _1 \kappa _3. \mbox{\hspace{1cm}}
\end{eqnarray*}

\vspace{-0.1cm}
Root structures of Adler-Moser polynomials are important to us, since we will link them to rogue wave patterns in the later text. Due to the free complex parameters $\{\kappa_j\}$ in them, their root structures will be understandably very diverse --- much more diverse than root structures of Yablonskii-Vorob'ev hierarchy polynomials. Indeed, when setting all $\{\kappa_j\}$ as zero except for one of them, we get root structures of Yablonskii-Vorob'ev hierarchy polynomials which are in the shape of triangles, pentagons, heptagons, and so on. When we continuously change those $\{\kappa_j\}$ values, we will get root structures which smoothly deform from one type of Yablonskii-Vorob'ev root structure to another, such as from a triangle to a pentagon. In this process, uncountably infinite new root shapes will be generated. These roots are generically simple roots. Indeed, if a root happens to be a multiple root, it will split into simple roots when the complex parameters $\{\kappa_j\}$ are slightly perturbed. For this reason, we will focus on the case when all roots of $\Theta_{N}(z)$ are simple in this article. In this case, $\Theta_{N}(z)$ will have $N(N+1)/2$ roots.

\begin{figure}[htb]
\begin{center}
\includegraphics[scale=0.28, bb=260 0 555 300]{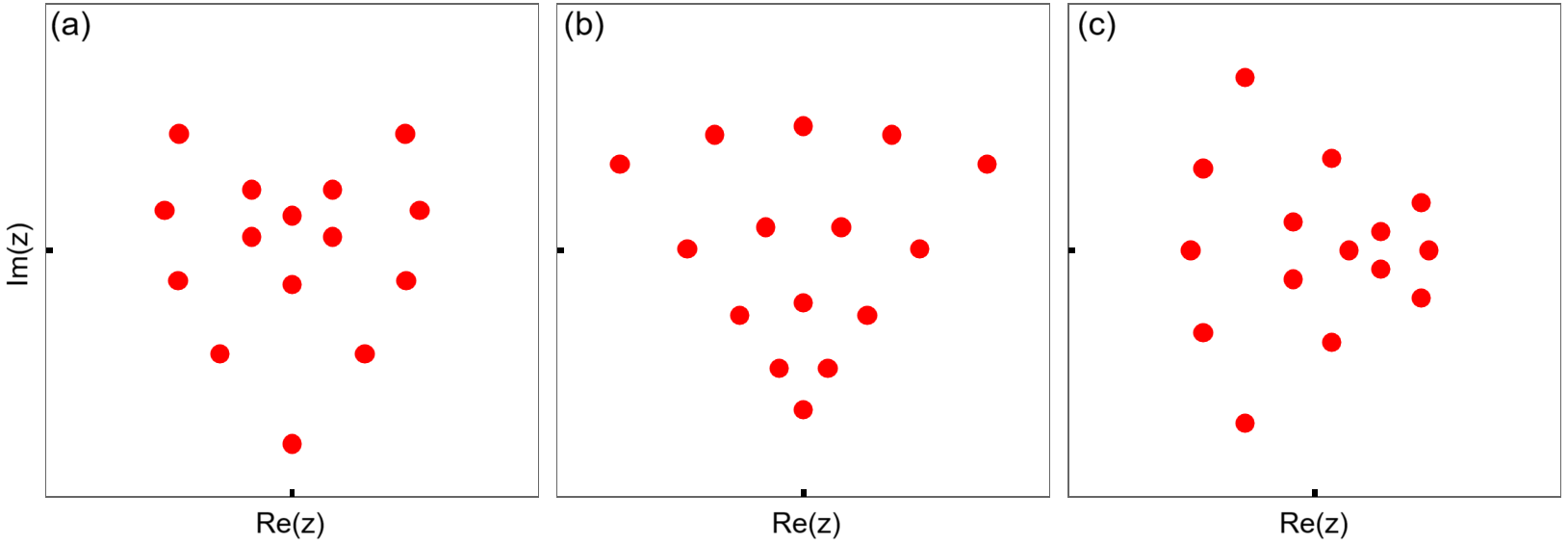}
\caption{Three sample root structures of $\Theta_{5}(z)$ for parameter values $(\kappa_1, \kappa_2, \kappa_3, \kappa_4)$ given in Eq.~(\ref{para}). In all panels, $-7\le \mbox{Re}(z), \mbox{Im}(z)\le 7$. } \label{f:roots}
\end{center}
\end{figure}

Of the uncountably infinite root structures of Adler-Moser polynomials, we illustrate only three of them for brevity. These three samples are for $\Theta_{5}(z; \kappa_1, \kappa_2, \kappa_3, \kappa_4)$, with three sets of $(\kappa_1, \kappa_2, \kappa_3, \kappa_4)$ values as
\[  \label{para}
(\textrm{i}, \textrm{i}, \textrm{i}, \textrm{i}), \hspace{0.1cm}
(5\textrm{i}/3, \textrm{i}, 5\textrm{i}/7, -5\textrm{i}/9), \hspace{0.1cm} (1, 1, 1, 1).
\]
Their root structures are displayed in Fig.~\ref{f:roots}(a, b, c), respectively. In these panels, every root is a simple root.
The (a) panel shows a heart-shaped structure, (b) shows a fan-shaped circular sector, and (c) shows a two-arc structure combined with a triangle.

\vspace{-1cm}
\section{Analytical predictions for rogue patterns with multiple large internal parameters}
Rogue wave solutions $u_N(x,t)$ in Eq.~(\ref{NLSNRWs}) contain $N-1$ free internal complex parameters $a_3, a_5, \cdots, a_{2N-1}$. If only one of those parameters is large, then the resulting rogue pattern is predicted by root structures of Yablonskii-Vorob'ev hierarchy polynomials, see \cite{YangNLS2021}. In this section, we consider patterns of these rogue solutions when \emph{multiple} of these internal parameters are large.

Specifically, suppose parameters $a_3, a_5, \cdots, a_{2N-1}$ in $u_N(x,t)$ are of the following form
\[ \label{acond}
a_{2j+1}=\kappa_j \hspace{0.04cm} A^{2j+1}, \quad 1\le j\le N-1,
\]
where $A\gg 1$ is a large positive constant, and $(\kappa_1, \kappa_2, \dots, \kappa_{N-1})$ are $O(1)$ complex constants not being all zero. Suppose also that roots of the Adler-Moser polynomial $\Theta_{N}(z)$ with parameters $\{\kappa_j\}$ are all simple. Then, our analytical prediction on the pattern of this rogue wave solution $u_N(x,t)$ is given by the following theorem.

\textbf{Theorem 1.} \emph{If all roots of $\Theta_{N}(z)$ are simple, then the $N$-th order rogue wave $u_N(x,t)$ in Eq.~(\ref{NLSNRWs}) with its internal large parameters $(a_3, a_5, \cdots, a_{2N-1})$ as given by Eq.~(\ref{acond}) would asymptotically split into $N(N+1)/2$ fundamental (Peregrine) rogue waves of the form $\hat{u}_1(x-\hat{x}_{0}, t-\hat{t}_{0}) \hspace{0.05cm} e^{\textrm{i}t}$, where $\hat{u}_1(x, t)$ is given in Eq.~(\ref{Pere}), and positions $(\hat{x}_{0}, \hat{t}_{0})$ of these Peregrine waves are given by}
\begin{equation}
\hat{x}_{0}+\textrm{i}\hspace{0.05cm}\hat{t}_{0}=z_{0}A, \label{x0t0}
\end{equation}
\emph{with $z_{0}$ being every one of the $N(N+1)/2$ simple roots of $\Theta_{N}(z)$. The error of this Peregrine wave approximation is $O(A^{-1})$. Expressed mathematically, when $(x-\hat{x}_{0})^2+(t-\hat{t}_{0})^2=O(1)$, we have the following solution asymptotics}
\begin{equation*}
u_{N}(x,t; a_{3}, a_{5}, \cdots, a_{2N-1}) = \hat{u}_1(x-\hat{x}_{0},t-\hat{t}_{0})\hspace{0.05cm} e^{\textrm{i}t} + O\left(A^{-1}\right).
\end{equation*}
\emph{When $(x,t)$ is not in the neighborhood of any of these Peregrine waves, $u_N(x,t)$ would asymptotically approach the constant-amplitude background $e^{\textrm{i}t}$ as $A\to +\infty$.}

This theorem indicates that the rogue pattern is asymptotically a simple dilation of the root structure of the underlying Adler-Moser polynomial by a factor of $A$, with each root predicting the location of a Peregrine wave in the $(x, t)$ plane according to Eq.~(\ref{x0t0}). Thus, this theorem establishes a direct connection between rogue patterns and root structures of Adler-Moser polynomials.

One may notice that in the present case of multiple large parameters, the rogue pattern is a simple dilation of the root structure of an Adler-Moser polynomial, while in the previous case of a single large parameter as studied in \cite{YangNLS2021}, the rogue pattern was a dilation \emph{and rotation} of the root structure of a Yablonskii-Vorob'ev hierarchy polynomial. The reason our current rogue pattern does not involve rotation to the root structure is that, the Adler-Moser polynomial contains free complex constants $\{\kappa_j\}$, which automatically put its root structure in proper orientation to match the rogue pattern. Comparatively, a Yablonskii-Vorob'ev hierarchy polynomial does not contain such free complex constants, and thus the orientation of its root structure is fixed. In this case, in order for its root structure to match the orientation of the rogue wave, a proper rotation is needed.

\vspace{-0.5cm}
\section{Numerical confirmation}

\vspace{-0.2cm}
Now, we numerically verify Theorem 1 by comparing its predictions with true rogue-wave solutions. This comparison will be done only for fifth-order rogue waves $u_5(x, t)$ for brevity. Such fifth-order solutions have internal complex parameters $(a_3, a_5, a_7, a_9)$.

We will do this comparison on three examples. Internal parameter values in these three examples are of the form (\ref{acond}) with $A=5$, which is large as desired, and their $(\kappa_1, \kappa_2, \kappa_3, \kappa_4)$ values are given in Eq.~(\ref{para}). These $\kappa_j$ values are used since root structures of Adler-Moser polynomials $\Theta_5(z)$ for these values have been displayed in Fig.~1. For these three sets of internal parameters, true rogue wave solutions are plotted in the upper three panels of Fig.~2, respectively. It is seen that each panel comprises 15 lumps (Peregrine waves) in the $(x, t)$ plane. In the first panel, these 15 Peregrine waves form a heart-shaped structure, with another mini-heart in its interior. In the second panel, these 15 Peregrine waves form a fan-shaped structure. In the third panel, these 15 Peregrine waves form two vertically-oriented arcs plus a smaller triangle on their right side.

\begin{figure}[htb]
\begin{center}
\includegraphics[scale=0.18, bb=950 0 300 890]{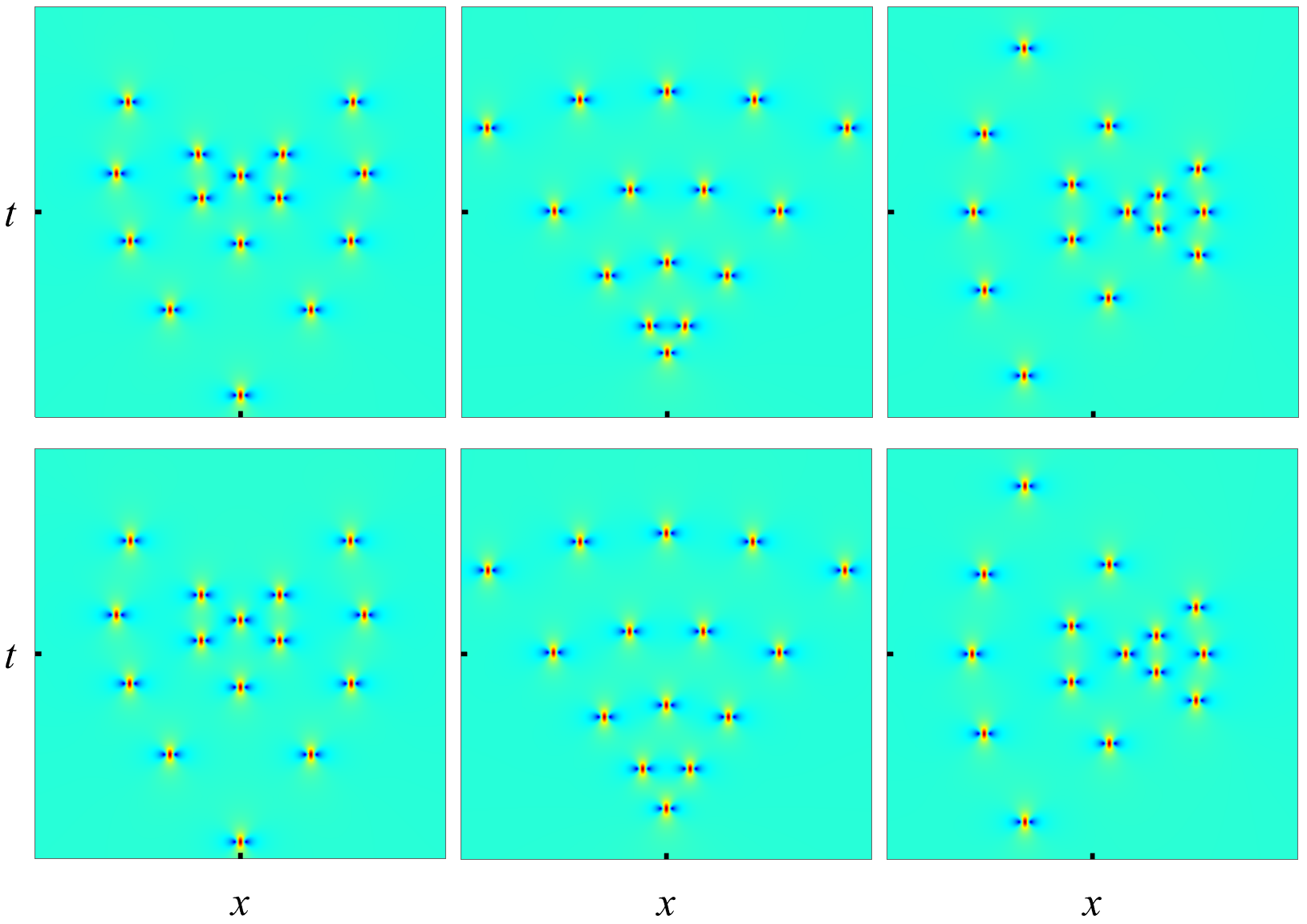}
\caption{Comparison between true rogue solutions $|u(x, t)|$ (upper row) and their analytical predictions (lower row) for $N=5$ and $A=5$. From left to right columns: $(\kappa_1, \kappa_2, \kappa_3, \kappa_4)=(\textrm{i}, \textrm{i}, \textrm{i}, \textrm{i}), \ (5\textrm{i}/3, \textrm{i}, 5\textrm{i}/7, -5\textrm{i}/9),\ (1, 1, 1, 1)$. In all panels, $-30 \le x, t\le 30$. }  \label{f:example2}
\end{center}
\end{figure}

Our analytical predictions $|u_5^{(p)}(x,t)|$ for these rogue waves from Theorem~1 can be assembled into a simple formula,
\begin{equation} \label{upNLS}
\left|u_5^{(p)}(x,t)\right|=1+\sum _{j=1}^{15}  \left(\left| \hat{u}_1(x-\hat{x}_{0}^{(j)}, t-\hat{t}_{0}^{(j)})\right| -1 \right),
\end{equation}
where $\hat{u}_1(x,t)$ is the Peregrine wave given in (\ref{Pere}), and their positions $(\hat{x}_{0}^{(j)}, \hat{t}_{0}^{(j)})$ are given by (\ref{x0t0}) with $z_0$ being every one of the $N(N+1)/2=15$ simple roots of the Adler-Moser polynomial $\Theta_5(z)$. These predicted solutions for the same $(a_3, a_5, a_7, a_9)$ values as in the true solutions are plotted in the lower three panels of Fig.~\ref{f:example2}. When compared to the root structures of the Adler-Moser polynomial $\Theta_5(z)$ in Fig.~\ref{f:roots}, our predicted rogue patterns in these lower panels are obviously a simple dilation of those root structures, by a factor of $A=5$, with each root replaced by a Peregrine wave, as Theorem~1 says.

When comparing the true rogue solutions in the upper row to their analytical predictions in the lower row, we can clearly see that they agree with each other very well. In fact, one can hardly notice the difference between them, which is an indication that our prediction in Theorem 1 is highly accurate.

Quantitatively, we have also measured the error of our analytical predictions versus the $A$ value, similar to what we did in Fig.~5 of Ref.~\cite{YangNLS2021}. That error analysis confirmed that the error does decay in proportion to $A^{-1}$, as Theorem~1 predicts. Thus, Theorem~1 is fully confirmed numerically. Details of this quantitative comparison are omitted here for brevity.

\section{Proof of Theorem 1} \label{sec:proof}

\vspace{-0.4cm}
In this section, we prove the analytical predictions on NLS rogue patterns in Theorems 1. The main idea of our proof resembles that in Ref.~\cite{YangNLS2021} for a single large internal parameter case.

To derive the large-parameter asymptotics of the rogue wave solution $u_N(x, t)$ in Eq.~(\ref{NLSNRWs}), we need asymptotic expressions for the determinant $\sigma_n$ in Eq.~(\ref{sigma_n}). For this purpose, we first use determinant identities and the Laplace expansion to rewrite $\sigma_n$ as \cite{OhtaJY2012}
\begin{eqnarray} \label{sigmanLap}
&& \hspace{-1.25cm} \sigma_{n}=\sum_{0\leq\nu_{1} < \nu_{2} < \cdots < \nu_{N}\leq 2N-1}
\det_{1 \leq i, j\leq N} \left[\frac{1}{2^{\nu_j}} S_{2i-1-\nu_j}(\textbf{\emph{x}}^{+}(n) +\nu_j \textbf{\emph{s}}) \right] \nonumber \\
&& \times \det_{1 \leq i, j\leq N}\left[\frac{1}{2^{\nu_j}}S_{2i-1-\nu_j}(\textbf{\emph{x}}^{-}(n) + \nu_j \textbf{\emph{s}})\right],
\end{eqnarray}
where $S_k\equiv 0$ if $k<0$.

When internal parameters $(a_3, a_5, \cdots, a_{2N-1})$ are of the form (\ref{acond}) with $A\gg 1$, and $x, t= O(A)$ or smaller, we have
\begin{eqnarray}\label{Skasym2}
&& \hspace{-0.7cm} S_{k}(\textbf{\emph{x}}^{+}(n) +\nu \textbf{\emph{s}}) =S_{k}\left(x_{1}^{+}, \nu s_{2}, x_{3}^{+}, \nu s_4, \cdots\right) \nonumber \\
&& \hspace{-0.7cm} =S_{k}\left(x_1^+, 0, \kappa_{1}A^{3}, 0, \kappa_{2}A^{5}, \cdots\right)\left[1+O(A^{-2})\right] \nonumber \\
&& \hspace{-0.7cm} = S_k(\hat{\textbf{v}}) \left[1+O(A^{-2})\right],  \label{Skxnasym2}
\end{eqnarray}
where $\hat{\textbf{v}}=\left( x+\textrm{i}t+n, 0, \kappa_1 A^3, 0, \kappa_2A^5, 0, \cdots\right)$.
From the definition (\ref{Schurdef}) of Schur polynomials, one can see that the polynomial $S_k(\hat{\textbf{v}})$ is related to $\theta_{k}(z)$ in (\ref{AMthetak}) as
\begin{equation} \label{Skorder2}
S_k(\hat{\textbf{v}})=A^{k} \theta_{k}(\hat{z}),
\end{equation}
where $\hat{z}\equiv A^{-1}(x+\textrm{i}t+n)$.

The dominant contribution in the Laplace expansion (\ref{sigmanLap}) of $\sigma_{n}$ comes from two index choices, $\nu=(0, 1, \cdots, N-1)$, and $\nu=(0, 1, \cdots, N-2, N)$.

With the first index choice, in view of Eqs. (\ref{Skxnasym2})-(\ref{Skorder2}), the determinant involving $\textbf{\emph{x}}^{+}(n)$ inside the summation of (\ref{sigmanLap}) is asymptotically
\[  \label{Phinnxtnew}
\alpha \hspace{0.06cm} A^{\frac{N(N+1)}{2}}\Theta_{N}(\hat{z}) \left[1+O\left(A^{-2}\right)\right],
\]
%\[ \label{Phinnxtnew}
%\det_{1 \leq i, j\leq N} \left[S_{2i-j}(\textbf{\emph{x}}^{+}(n) +j \textbf{\emph{s}}) \right]
%\sim \alpha \hspace{0.06cm} A^{\frac{N(N+1)}{2}}\Theta_{N}(\hat{z}) \left[1+O\left(A^{-2}\right)\right],
%\]
where $\alpha=2^{-N(N-1)/2}c_N^{-1}$. Let us define $(\hat{x}_{0}, \hat{t}_{0})$ by Eq.~(\ref{x0t0}), i.e.,
$z_0=A^{-1}(\hat{x}_{0}+\textrm{i}\hat{t}_{0})$, where $z_0$ is a simple root of the Adler-Moser polynomial $\Theta_{N}(z)$. Then, when $(x, t)$ is in the $O(1)$ neighborhood of $(\hat{x}_{0}, \hat{t}_{0})$, we expand $\Theta_{N}(\hat{z})$ around $\hat{z}=z_0$.  Recalling $\Theta_{N}(z_0)=0$, we get
\begin{equation*}
\hspace{-0.5cm}\Theta_{N}(\hat{z})=A^{-1}\left[(x-\hat{x}_{0})+\textrm{i}(t-\hat{t}_{0})+n\right] \Theta'_{N}(z_0)
\left[1+O\left(A^{-1}\right)\right].
\end{equation*}
Inserting this equation into (\ref{Phinnxtnew}), the determinant involving $\textbf{\emph{x}}^{+}(n)$ inside the summation of (\ref{sigmanLap}) becomes
\begin{equation*}
\left[(x-\hat{x}_{0})+\textrm{i}(t-\hat{t}_{0})+n\right] \hspace{0.06cm} \alpha\hspace{0.06cm} A^{\frac{N(N+1)}{2}}\Theta_{N}'(z_0)\left[1+O\left(A^{-1}\right)\right].
\end{equation*}
Similarly, the determinant involving $\textbf{\emph{x}}^{-}(n)$ inside this summation becomes
\begin{equation*}
\left[(x-\hat{x}_{0})-\textrm{i}(t-\hat{t}_{0})-n\right] \hspace{0.06cm} \alpha \hspace{0.06cm} A^{\frac{N(N+1)}{2}}\Theta_{N}'(z_0^*)\left[1+O\left(A^{-1}\right)\right].
\end{equation*}

Next, we consider the contribution from the second index choice of $\nu=(0, 1, \cdots, N-2, N)$. For this index choice, the determinant involving $\textbf{\emph{x}}^{+}(n)$ inside the summation of (\ref{sigmanLap}) becomes
\begin{equation*}
\frac{1}{2}\alpha \hspace{0.06cm} A^{\frac{N(N+1)-2}{2}}\Theta_{N}'(\hat{z}) \left[1+O\left(A^{-2}\right)\right].
\end{equation*}
When $(x, t)$ is in the $O(1)$ neighborhood of $(\hat{x}_{0}, \hat{t}_{0})$, the above term is asymptotically equal to
\begin{equation*}
\frac{1}{2} \alpha \hspace{0.06cm} A^{\frac{N(N+1)-2}{2}}\Theta_{N}'(z_0) \left[1+O\left(A^{-1}\right)\right].
\end{equation*}
Similarly, the determinant involving $\textbf{\emph{x}}^{-}(n)$ inside the summation of (\ref{sigmanLap}) becomes
\begin{equation*}
\frac{1}{2} \alpha \hspace{0.06cm} A^{\frac{N(N+1)-2}{2}}\Theta_{N}'(z_0^*) \left[1+O\left(A^{-1}\right)\right].
\end{equation*}

Summarizing the above two dominant contributions in the Laplace expansion (\ref{sigmanLap}), we find that
\begin{eqnarray}  \label{sigmanxt5}
&& \hspace{-1cm} \sigma_{n}(x,t) = \alpha^2 \hspace{0.06cm} \left|\Theta_{N}'(z_0)\right|^2 A^{N(N+1)-2} \nonumber \\
&& \hspace{-0.5cm} \times \left[ \left(x-\hat{x}_{0}\right)^2+\left(t-\hat{t}_{0}\right)^2 - 2 \textrm{i} n \left(t-\hat{t}_{0}\right)-n^2+\frac{1}{4} \right] \nonumber \\
&& \hspace{-0.5cm} \times \left[1+O\left(A^{-1}\right)\right].
\end{eqnarray}
Since the root $z_0$ has been assumed simple, $\Theta_{N}'(z_0)\ne 0$. Thus, the above leading-order asymptotics for $\sigma_{n}(x,t)$ does not vanish. Therefore, when $A$ is large and $(x, t)$ in the O(1) neighborhood of $\left(\hat{x}_{0}, \hat{t}_{0}\right)$, we get from (\ref{sigmanxt5}) that
\begin{eqnarray*}
&& \hspace{-1.2cm} u_N(x,t) = \frac{\sigma_{1}}{\sigma_{0}}e^{\textrm{i}t} =e^{\textrm{i}t}
\left(1- \frac{4[1+2\textrm{i}(t-\hat{t}_{0})]}{1+4(x-\hat{x}_{0})^2+4(t-\hat{t}_{0})^2}\right) \nonumber \\
&& \hspace{1.4cm} + O\left(A^{-1}\right),
\end{eqnarray*}
which is a Peregrine wave $\hat{u}_1(x-\hat{x}_{0}, t-\hat{t}_{0}) \hspace{0.05cm} e^{\textrm{i}t}$, and the error of this Peregrine prediction is $O\left(A^{-1}\right)$. Theorem~1 is then proved.

\vspace{0.5cm}

\section{Conclusions and discussions}
In this paper, we have reported many new rogue patterns in the NLS equation which are predicted by root structures of new special polynomials. Specifically, we have shown that when multiple internal parameters in the rogue wave solutions are large, many new rogue patterns would arise, including heart-shaped structures, fan-shaped structures, and others. Analytically, these rogue patterns are determined by root structures of Adler-Moser polynomials. If all roots of the Adler-Moser polynomial are simple, then the rogue pattern is simply a dilation of the Adler-Moser-polynomial's root structure, with each root replaced by a Peregrine wave. Since Adler-Moser polynomials contain free complex parameters, their root structures would be very diverse. As a result, NLS rogue waves could assume much more varied spatial-temporal patterns beyond those reported earlier.

Since NLS rogue waves have been observed in diverse physical systems \cite{Kibler2010,Xu2019,Chabchoub2011,Chabchoub2012a,Chabchoub2012b,Chabchoub2013,Helium2008,Bailung2011,Engles2023}, these new NLS rogue patterns open up more varieties of rogue dynamics which could be verified in experiments too.

The previous NLS rogue patterns associated with root structures of the Yablonskii-Vorob'ev polynomial hierarchy were later found to be universal and would appear in many other integrable systems \cite{YangNLS2021,YangNLS2021b}. The present rogue patterns associated with Adler-Moser polynomials are expected to be universal as well, and they should arise in other integrable systems too when multiple internal parameters in rogue waves of those integrable equations are large. This prospect will be pursued in the near future.

Our rogue-pattern predictions in this article were made under the assumption that all roots of the Adler-Moser polynomial are simple. A very interesting question is what will happen if some roots of the Adler-Moser polynomial are not simple. This question is still open that merits further studies.

\vspace{-0.5cm}
\section*{Acknowledgment}
The work of B.Y. was supported in part by the National Natural Science Foundation of China (GrantNo.12201326), and the work of J.Y. was supported in part by the National Science Foundation (U.S.) under award number DMS-1910282.


\vspace{-0.5cm}
\section*{References}

\begin{thebibliography}{10}
\bibitem{Peli_book2009}
C. Kharif, E. Pelinovsky, and A. Slunyaev, \emph{Rogue Waves in the Ocean} (Springer, Berlin, 2009).
\bibitem{Solli2007}
D.R. Solli, C. Ropers, P. Koonath, and B. Jalali,
%``Optical rogue waves",
Nature 450, 1054 (2007).
\bibitem{Wabnitz2017}
S. Wabnitz (ed.), \emph{Nonlinear Guided Wave Optics: A Testbed for Extreme Waves} (IOP Publishing, Bristol, 2017).
\bibitem{RomeroRos2022}
A. Romero-Ros, G. C. Katsimiga, S. I. Mistakidis, B. Prinari, G. Biondini, P. Schmelcher, and
P. G. Kevrekidis,
%``Theoretical and numerical evidence for the potential realization of the Peregrine soliton in repulsive two-component Bose-Einstein condensates",
Phys. Rev. A 105, 053306 (2022).
\bibitem{Kibler2010}
B. Kibler, J. Fatome, C. Finot, G. Millot, F. Dias, G. Genty, N. Akhmediev and J. M. Dudley,
%``The Peregrine soliton in nonlinear fibre optics",
Nature Phys. 6, 790 (2010).
\bibitem{Xu2019}
G. Xu, K. Hammani, A. Chabchoub, J. M. Dudley, B. Kibler, and C. Finot,
%``Phase evolution of Peregrine-like breathers in optics and hydrodynamics",
Phys. Rev. E 99, 012207 (2019).
\bibitem{Frisquet2016}
B. Frisquet, B. Kibler, P. Morin, F. Baronio, M. Conforti, G. Millot, and S. Wabnitz,
%``Optical dark rogue waves",
Sci. Rep. 6, 20785 (2016).
\bibitem{Baronio2018}
F. Baronio, B. Frisquet, S. Chen, G. Millot, S.Wabnitz, and B. Kibler,
%``Observation of a group of dark rogue waves in a telecommunication optical fiber",
Phys. Rev. A 97, 013852 (2018).
\bibitem{Chabchoub2011}
A. Chabchoub, N.P. Hoffmann, and N. Akhmediev,
%``Rogue wave observation in a water wave tank",
Phys. Rev. Lett. 106, 204502 (2011).
\bibitem{Chabchoub2012a}
A. Chabchoub, N. Hoffmann, M. Onorato, and N. Akhmediev,
%``Super rogue waves: observation of a higher-order breather in water waves",
Phys. Rev. X 2, 011015 (2012).
\bibitem{Chabchoub2012b}
A. Chabchoub, N. Hoffmann, M. Onorato, A. Slunyaev, A. Sergeeva, E. Pelinovsky and N.
Akhmediev,
%``Observation of a hierarchy of up to fifth-order rogue waves in a water tank",
Phys. Rev. E 86, 056601 (2012).
\bibitem{Chabchoub2013}
A. Chabchoub and N. Akhmediev,
%``Observation of rogue wave triplets in water waves",
Phys. Lett. A 377, 2590 (2013).
\bibitem{Helium2008}
A.N. Ganshin, V.B. Efimov, G.V. Kolmakov, L.P. Mezhov-Deglin and P.V.E. McClintock,
%``Observation of an inverse energy cascade in developed acoustic turbulence in superfluid helium",
Phys. Rev. Lett. 101, 065303 (2008).
\bibitem{Bailung2011}
H. Bailung, S.K. Sharma and Y. Nakamura,
%``Observation of Peregrine solitons in a multicomponent plasma with negative ions",
Phys. Rev. Lett. 107, 255005 (2011).
\bibitem{Plasma2016}
Y.Y. Tsai, J.Y. Tsai and L. I,
%``Generation of acoustic rogue waves in dusty plasmas through three-dimensional particle focusing by distorted waveforms",
Nat. Phys. 12, 573 (2016).
\bibitem{Engles2023}
A. Romero-Ros, G.C. Katsimiga, S.I. Mistakidis, S. Mossman, G. Biondini, P. Schmelcher,
P. Engels, and P.G. Kevrekidis,
%``Experimental realization of the Peregrine soliton in repulsive two-component Bose-Einstein condensates",
arXiv:2304.05951 [nlin.PS] (2023).
\bibitem{Benney}
D.J. Benney and A.C. Newell,
%``The propagation of nonlinear wave envelopes",
J. Math. Phys. 46, 133 (1967).
\bibitem{Ablowitzbook}
M.J. Ablowitz and H. Segur, \emph{Solitons and the Inverse Scattering Transform} (SIAM, Philadelphia, 1981).
\bibitem{Hasegawabook}
A. Hasegawa and Y. Kodama, \emph{Solitons in Optical Communications} (Clarendon Press, Oxford, 1995).
\bibitem{Saito1984}
M. Saito, S. Watanabe and H. Tanaca,
%``Modulation instability of ion wave in plasma with negative ion",
J. Phys. Soc. Japan 53, 2304 (1984).
\bibitem{Gross}
E.P. Gross,
%``Structure of a quantized vortex in boson systems".
Il Nuovo Cimento. 20, 454 (1961).
\bibitem{Pitaevskii}
L.P. Pitaevskii,
%``Vortex lines in an imperfect Bose gas",
Sov. Phys. JETP. 13, 451 (1961).
\bibitem{Kevrekidis2022}
A. Romero-Ros, G. C. Katsimiga, S. I. Mistakidis, B. Prinari, G. Biondini, P. Schmelcher, and
P. G. Kevrekidis,
%``Theoretical and numerical evidence for the potential realization of the Peregrine soliton in repulsive two-component Bose-Einstein condensates ,
Phys. Rev. A 105, 053306 (2022).
\bibitem{Bergano}
S.G. Evangelides, L.F. Mollenauer, J.P. Gordon, and N.S. Bergano,
%``Polarization multiplexing with solitons",
J. Lightwave Technol. 10, 28 (1992).
\bibitem{Peregrine1983}
D.H. Peregrine,
%``Water waves, nonlinear Schr\"dinger equations and their solutions",
J. Aust. Math. Soc. B 25, 16 (1983).
\bibitem{AAS2009}
N. Akhmediev, A. Ankiewicz and J.M. Soto-Crespo,
%``Rogue waves and rational solutions of the nonlinear Schr\"odinger equation,"
Phys. Rev. E 80, 026601 (2009).
\bibitem{DGKM2010}
P. Dubard, P. Gaillard, C. Klein and V.B. Matveev,
%``On multi-rogue wave solutions of the NLS equation and positon solutions of the KdV equation,"
Eur. Phys. J. Spec. Top. 185, 247 (2010).

\bibitem{AKA2011}
A. Ankiewicz, D. Kedziora and N. Akhmediev,
%``Rogue wave triplet",
Phys. Lett. A 375, 2782 (2011).

\bibitem{KAAN2011}
D.J. Kedziora, A. Ankiewicz and N. Akhmediev,
%``Circular rogue wave clusters,"
Phys. Rev. E  84, 056611 (2011).

\bibitem{GLML2012}
B.L. Guo, L.M. Ling and Q.P. Liu,
%``Nonlinear Schr\"odinger equation: generalized Darboux transformation and rogue wave solutions,"
Phys. Rev. E 85, 026607 (2012).

\bibitem{OhtaJY2012}
Y. Ohta and J. Yang,
%``General high-order rogue waves and their dynamics in the nonlinear Schr\"odinger equation,"
Proc. R. Soc. Lond. A 468, 1716 (2012).

\bibitem{Miller2019}
D. Bilman and P.D. Miller,
%``A robust inverse scattering transform for the focusing nonlinear Schr\"odinger equation",
Commun. Pure Appl. Math. 72, 1722 (2019).

\bibitem{BDCW2012}
F. Baronio, A. Degasperis, M. Conforti and S. Wabnitz,
%``Solutions of the vector nonlinear Schr\"odinger equations: evidence for deterministic rogue waves",
Phys. Rev. Lett. 109, 044102 (2012).

\bibitem{ManakovDark}
F. Baronio, M. Conforti, A. Degasperis, S. Lombardo, M. Onorato and S. Wabnitz,
%``Vector rogue waves and baseband modulation instability in the defocusing regime",
Phys. Rev. Lett. 113, 034101 (2014).

\bibitem{LingGuoZhaoCNLS2014}
L. Ling, B. Guo and L. Zhao,
%``High-order rogue waves in vector nonlinear Schr\"{o}dinger equations",
Phys. Rev. E 89, 041201(R) (2014).

\bibitem{Chen_Shihua2014}
S. Chen, J. M. Soto-Crespo, and P. Grelu,
%``Dark three-sister rogue waves in normally dispersive optical fibers with random birefringence",
Opt. Express 22, 27632 (2014).

\bibitem{Chen_Shihua2015}
S. Chen and D. Mihalache,
%``Vector rogue waves in the Manakov system: diversity and compossibility",
J. Phys. A 48, 215202 (2015).

\bibitem{ZhaoGuoLingCNLS2016}
L. Zhao, B. Guo and L. Ling,
%``High-order rogue wave solutions for the coupled nonlinear Schr\"odinger equations-II",
J. Math. Phys. 57, 043508 (2016).

\bibitem{HeFokas}
J.S. He, H.R. Zhang, L.H. Wang, K. Porsezian and A.S. Fokas,
%``Generating mechanism for higher-order rogue waves",
Phys. Rev. E 87, 052914 (2013).

\bibitem{KAAN2013}
D.J. Kedziora, A. Ankiewicz and N. Akhmediev,
%``Classifying the hierarchy of nonlinear-Schr\"{o}dinger-equation rogue-wave solutions."
Phys. Rev. E 88, 013207 (2013).

\bibitem{YangNLS2021} B. Yang and J. Yang,
%``Rogue wave patterns in the nonlinear Schrodinger equation",
Physica D 419, 132850 (2021).

\bibitem{YangNLS2021b}
B. Yang and J. Yang,
%``Universal rogue wave patterns associated with the Yablonskii-Vorob ev polynomial hierarchy",
Physica D 425, 132958 (2021).

\bibitem{Adler_Moser1978}
M. Adler and J. Moser,
%``On a class of polynomials associated with the Korteweg de Vries equation",
Commun. Math. Phys. 61, 1 (1978).

\bibitem{Aref2007FDR}
H. Aref,
%``Vortices and polynomials",
Fluid Dynam. Res. 39, 5 (2007).

\bibitem{Clarkson2009}
P.A. Clarkson,
%``Vortices and polynomials",
Stud. Appl. Math. 123, 37 (2009).

\end{thebibliography}
\end{document}